# Atomic structure of a single large biomolecule from diffraction patterns of random orientations

Miklós Tegze* and Gábor Bortel

*Institute for Solid State Physics and Optics, Wigner Research Centre for Physics, Hungarian Academy of Sciences, H-1525 Budapest, P. O. Box 49, Hungary*

**Abstract**

The short and intense pulses of the new X-ray free electron lasers, now operational or under construction, may make possible diffraction experiments on single molecule-sized objects with high resolution, before radiation damage destroys the sample. In a single molecule imaging (SMI) experiment thousands of diffraction patterns of single molecules with random orientations are recorded. One of the most challenging problems of SMI is how to assemble these noisy patterns of unknown orientations into a consistent single set of diffraction data. Here we present a new method which can solve the orientation problem of SMI efficiently even for large biological molecules and in the presence of noise. We show on simulated diffraction patterns of a large protein molecule, how the orientations of the patterns can be found and the structure to atomic resolution can be solved. The concept of our algorithm could be also applied to experiments where images of an object are recorded in unknown orientations and/or positions like in cryoEM or tomography.

*Corresponding author. e-mail: mt@szfki.hu, tel: +36-1-3922222, fax:+36-1-3922219

**Introduction**

In the hundred years since the discovery of the X-rays, diffraction on crystals was the basic method of structure determination at atomic level. Developments in the last decades (Chapman and Nugent, 2010) aimed to recover the phase of the scattered beam, but they still require periodicity of the object or a beam much smaller than the sample (Thibault et al. 2008). Sub-nanometre imaging of molecule-sized objects became only recently possible with the emergence of single particle electron cryo-microscopy (cryoEM, van Heel et al., 2000). The latest development in the field was the proposed use of femtosecond pulses of an X-ray free-electron laser (XFEL) for single molecule imaging (SMI, Neutze et al., 2000, Huldt et al., 2003).

In a single molecule imaging experiment, identical molecules are injected repeatedly into an intense XFEL beam and their diffraction pattern is recorded. The basic idea behind this type of experiment is that the time of the measurement (the length of the XFEL pulse) is shorter than the time necessary for the radiation damage to develop (Neutze et al., 2000). The motion of atoms and ions during the pulse has been studied numerically, showing that atomic displacement is negligible in the first 5 fs of the pulse (Jurek et al. 2004a, 2004b, Jurek and Faigel 2008, Hau-Riege et al., 2005). The feasibility of the experiment has been demonstrated on larger objects and at less than atomic resolution (Chapman et al. 2006, 2011, Loh et al. 2010, Seibert et al. 2011).

In an SMI experiment a great number of diffraction patterns of single molecules are recorded. The patterns correspond to molecules with random orientations and are extremely noisy. These noisy patterns of unknown orientations have to be assembled into a consistent single set of diffraction data. Here we present a new method which can solve the orientation problem of SMI efficiently even for large biological molecules and in the presence of noise. We show on the example of a large protein molecule, that 100,000 diffraction patterns are sufficient to solve the orientation problem and the structure, less than estimated in earlier studies (Jurek et al., 2004, Fung et al., 2009). Although we developed this algorithm specifically for SMI, its concept could be applied to experiments where images of an object are recorded in unknown orientations and/or positions like in cryoEM or tomography.

The ability to solve the structure from the assembled three-dimensional data set has also been demonstrated (Fung et al. 2009, Loh and Elser 2009, Loh et al, 2010, Miao et al. 1999, 2001). A large (~100 kDa) biological molecule scatters only about $10^{-2}$ photons into a Shannon-Nyquist pixel (Huldt et al., 2003) at 1.9 Å resolution out of the estimated $4 \cdot 10^{12}$ photons of a single pulse of a next-generation XFEL source. Earlier studies (Jurek et al. 2004b) estimate that at least $2 \cdot 10^6$ randomly oriented diffraction patterns are needed for classification, orientation, assembly and successful solution of the structure of such a molecule. We will show here that $10^5$ diffraction patterns are sufficient to solve first the orientation problem and then the structure to atomic resolution.

Diffraction patterns represent spherical sections of a three-dimensional (3D) intensity distribution. To orient them relative to each other, two properties can be used: 1. Two patterns oriented differently always contain a common line along their intersection where they are identical, not regarding the noise. 2. Patterns with close orientations are similar. Orientation methods can use one of the above properties or both. Common line methods (Shneerson et al. 2008, Bortel and Tegze 2011), first developed for electron microscopy (van Heel et al. 2000), use the first property only. Since in these methods only a small part of the patterns are used for the determination of their relative orientation, they are more sensitive to noise. Recently, two iterative methods were developed which use both properties of the patterns. One is the expectation maximization (EMC) algorithm of Loh and Elser (2009), the other is the method by Fung et al. (2009) applying the generative topographic mapping (GTM) algorithm (Svensen, 1998). Both methods can successfully operate on remarkably noisy diffraction patterns. It was shown later (Moths and Ourmazd, 2011) that they represent two implementations of the same fundamental approach. They also share the same drawback: computationally they are not very efficient (Moths and Ourmazd, 2011).

An efficient iterative orientation method should be suitably simple to reduce calculation time, converge quickly to the true solution and give accurate orientations for the individual measured patterns. It should also tolerate the high noise level in the patterns. In this paper we present a new orientation method which satisfies the above criteria. In the following sections we describe the details of the algorithm and show on simulated diffraction patterns of a large protein molecule how the orientation problem and the structure can be solved.

**Orientation method**

The basic idea of our method is the following (Fig. 1). If we knew an approximate three-dimensional intensity distribution, we could compare an individual diffraction pattern to this distribution in all possible orientations (on an orientation grid) and find which orientation gives the best fit. After determining the best-fitting orientations for all patterns, we can construct a new, improved 3D intensity distribution from the patterns. Repeating these steps, we can hope that the procedure converges to the true orientations and the true 3D intensity. Unfortunately, we do not know an approximate 3D intensity distribution, but we can start from a random one. It is clear, that the true solution is a fixed point of the algorithm. However, it is not obvious that starting from an arbitrary distribution the iteration process converges to this solution or converges at all. Although we cannot prove the convergence of the method analytically, we will show on numerical examples that the method converges very fast to the true solution. Our method can be regarded as a much simplified version of the EMC method of Loh and Elser (2009), which inspired it. In our algorithm the rather complicated and computationally demanding step of the maximization of the intensity's likelihood function is replaced by a simple search for the best-fitting orientation. This simplification also allows the use of efficient numerical methods which improve the scaling properties of the algorithm.

*Orientation grid*

The 3D scattering intensity distribution is the square of the absolute value of the Fourier transform of the electron density of the object. When the object rotates in real space, this intensity distribution rotates together with the object in reciprocal space. A single diffraction pattern can be regarded as a spherical cap (part of the Ewald-sphere) cut from this 3D intensity distribution. The centre of the pattern is at the origin of the reciprocal space and the pattern has a spherical curvature with radius $k=2\pi/\lambda$. In order to find the best-fitting orientation of a measured diffraction pattern, we have to cut out spherical sections in all possible orientations from the 3D intensity distribution. For this we use a discrete grid in the space of orientations.

Orientation of an object can be given by the three Euler angles. There are many definitions of Euler angles; here we define them with rotations about the fixed (laboratory) coordinate axes. Starting from a fixed reference orientation, orientation of the object is described by subsequent rotations about axes *z*, *x*, and *z* by Euler-angles $\Phi$, $\Theta$, and $\Psi$, respectively. If the incident beam is unpolarized (if not, the diffraction patterns can be corrected by the polarization factor), the diffraction pattern does not change when the object is rotated about the incident beam (z-axis) but simply rotates together with the object. Therefore, the angle $\Psi$ can and will be treated differently from $\Theta$ and $\Phi$. For the latter two angles, we set up an approximately uniform grid of the ($\Theta,\Phi$) subspace by dividing the faces of an icosahedron to smaller identical triangles and projecting them to the surface of a unit sphere. A similar construction for the 3D orientation space was introduced by Loh & Elser (2009). If the edge of the icosahedron is divided into *n* equidistant parts, then the number of points in the grid is $N_R = 10 n^2 + 2$. We improved the uniformity of the grid by slightly modifying the sizes of the triangles so, that the projection of the edge of the icosahedron has equidistant division, instead of the edge itself. The grid points on the sphere have spherical coordinates $\Theta_r$ and $\Phi_r$, $r=1\ldots N_R$. Each of the $N_R$ points of this orientation grid corresponds to a rotation matrix $\mathbf{R}_r$. We note here, that when the measured patterns are rotated relative to the 3D intensity distribution fixed to the object, we have to use the inverse of this rotation matrix $\mathbf{R}_r^T$.

*Data pre-processing*

The diffraction patterns are taken as an intensity distribution (photon counts) on a polar grid:

$$M_{mij} = M_m(\mathbf{q}_{ij}) = M_m(\mathbf{q}(\vartheta_i, \varphi_j)), \quad i=1\ldots N_\vartheta, \ j=1\ldots N_\varphi, \ m=1\ldots N_M . \tag{1}$$

Here **q** is the scattering vector and $N_M$ is the number of measured patterns. The polar grid is uniform in both the azimuthal ($\varphi$) and polar ($\vartheta$) directions with steps $\Delta\varphi$ and $\Delta\vartheta \approx \Delta\varphi \sin\vartheta_{max}$, respectively. We remove the strong radial dependence of the intensity common in all pictures (dividing by $\frac{1}{N_M N_\varphi} \sum_{m,j} M_{mij}$) and apply a low-pass filter by convolving the patterns with a Gaussian of FWHM = 2 pixels. The resulting $\tilde{M}_{mij}$ patterns are used in the iteration process.

In real SMI experiments the diffraction patterns are recorded on flat two-dimensional detectors with a Cartesian grid. In that case an extra preparatory step is needed: The detector patterns first have to be transformed to the polar grid described above. This could be done by some kind of interpolation and/or averaging. This processing may introduce some smoothing of the pattern, which may replace the application of the low-pass filter.

*Details of the algorithm*

A chart of the algorithm is shown in Fig. 1. We start the iteration with a random intensity distribution $I(\mathbf{q}_s)$. The vectors $\mathbf{q}_s$ define a three-dimensional Cartesian grid of $N_I$ points (voxels) in the reciprocal space. We sample this distribution by cutting out $I_{rij}=I(\mathbf{R}_r^T\mathbf{q}_{ij})$ spherical caps (parts of the Ewald-sphere), corresponding to diffraction patterns of the particle in different orientations $r$ of the approximately uniform 2D orientation grid. Since the rotated vectors $\mathbf{R}_r^T\mathbf{q}_{ij}$ describing these spherical sections do not coincide with the vectors $\mathbf{q}_s$ of the 3D grid, we choose the intensity corresponding to the nearest $\mathbf{q}_s$ vector. This correspondence between indices $rij$ and $s$ is tabulated in a matrix $s_{rij}$. Thus the transformation from the 3D intensity distribution to $N_R$ patterns of different orientations can be written as $I_{rij} = I(\mathbf{q}_{s_{rij}})$.

The similarity between two sets of data $\{a_j\}$ and $\{b_j\}$, $j = 1\ldots N$, can be expressed in terms of the Pearson correlation (Rodgers and Nicewander, 1988):

$$C_P\{a_j,b_j\}=\frac{\sum_j (a_j-\bar{a})(b_j-\bar{b})}{\sqrt{\sum_j (a_j-\bar{a})^2}\sqrt{\sum_j (b_j-\bar{b})^2}}, \text{ where } \bar{a}=\frac{1}{N}\sum_{j=1}^{N}a_j. \qquad (2)$$

We compute the correlation between all possible pairs of measured patterns $\widetilde{M}_{mij}$ and spherical sections $I_{rij}$ cut from the 3D intensity distribution in all possible relative rotations $\mathbf{R}_{\varphi_{j'}}$ about the centre (equivalent to rotation about the z-axis by $\Psi=\varphi_{j'}$):

$$c_{mrj'}=\frac{1}{N_\vartheta}\sum_i C_P\left\{\widetilde{M}_{mij}, I\left(\mathbf{R}_r^T\mathbf{R}_{\varphi_{j'}}\mathbf{q}_{ij}\right)\right\} \qquad (3)$$

Here we calculated the Pearson correlations between pairs of the corresponding resolution circles (circles of constant $\vartheta$ values) and averaged these correlation values for all circles.

We sample the $\mathbf{R}_{\varphi_{j'}}$ rotations about the centre of the pattern with the same $\Delta\varphi$ step as the azimuthal one of the polar grid of the pattern. In this case, the $\mathbf{R}_{\varphi_{j'}}$ rotation is the same as a shift in the index j of the reciprocal space vector:

$$\mathbf{R}_{\varphi_{j'}}\mathbf{q}_{ij}=\mathbf{q}_{ij-j'} \qquad (4)$$

Applying (4) in expression (3), the correlation becomes:

$$c_{mrj'}=\frac{1}{N_\vartheta}\sum_i C_P\left\{\widetilde{M}_{mij}, I_{rij-j'}\right\} \qquad (5)$$

The $c_{mrj'}$ correlation matrix elements can be calculated for all $\varphi_{j'}$ angles at the same time with the help of fast Fourier transform (FFT) algorithms using the cross-correlation theorem (Weisstein, 2012). Now we find the orientation indices $r_m$ and $j_m$ where the correlation is at maximum:

$$c_m^{max} = c_{mr_m j_m} = \max_{r,j'}\{c_{mrj'}\} \tag{6}$$

To increase efficiency, the calculation of the correlation matrix elements and their maxima was implemented for graphics processors using the CUDA software package (CUDA Zone, 2012).

We use the orientations $(r_m, j_m)$ corresponding to the correlation maxima to rotate the patterns to their most fitting positions, multiply them by the scaling factor $a_m = \sum_{ij} I_{r_m ij} / \sum_{ij} \tilde{M}_{mij}$ and construct a new 3D intensity distribution by averaging the contributions of points nearest to the $\mathbf{q}_s$ points of the 3D grid for all oriented patterns. The corresponding voxels are selected using the tabulated $s_{rij}$ matrix elements with $r=r_m$. We found that convergence is faster if only those patterns contribute to the new intensity distribution which have $c_m^{max} > \min(\text{median}(c_m^{max}), 0.1)$.

When the new intensity distribution is constructed, we start a new iteration. Convergence of the algorithm (i.e. finding the proper relative orientation of the patterns) should appear as a pronounced increase in the maximum correlation values $c_m^{max}$. After convergence, the orientation angles are refined using a simplex search method for finding the angles of maximum correlation. We can use these refined orientations to construct a 3D intensity distribution in reciprocal space from the original $M_m$ measured patterns for structure reconstruction.

*Scaling properties of the algorithm*

An important property of an algorithm is how it scales with the size of the problem. The single parameter governing the difficulty of the task is the ratio $R=D/d$ between the diameter of the particle $D$ and the spatial resolution $d$. The $N_M$ randomly oriented diffraction patterns represent $N_M \times N_\vartheta \times N_\varphi$ data points distributed inside a sphere in the reciprocal space. The necessary number of grid points $N_I$ in the 3D reciprocal space scales as $R^3$ (Loh and Elser, 2009). Obviously, the $q$-spacing of the points in the 2D patterns should match the spacing of the grid points in the reciprocal space. Therefore, both $N_\vartheta$ and $N_\varphi$ scales with $R$. Following the arguments of Loh and Elser (2009), we find that the necessary number of points $N_R$ in the $(\Theta,\Phi)$ orientational subspace scales with $R^2$. Now we have to estimate the number of measurements $N_M$ needed for the orientation algorithm to work. They should give contribution to all $N_I$ grid points of the 3D reciprocal space, so $N_M$ scales as $N_I/(N_\vartheta \times N_\varphi) \propto R$. Since the data points are not evenly distributed and the data are noisy, the necessary number could be much larger. However, we can suppose that this factor is independent of $R$ and does not change the conclusion.

The most time-consuming part is the calculation of the correlation matrix, having $N_\varphi \times N_R \times N_M$ elements. The time needed to calculate a single matrix element is proportional to the number of data points in the pattern $N_\vartheta \times N_\varphi$. If calculated sequentially, the calculation time of the full correlation matrix would be proportional to $N_\vartheta \times N_\varphi^2 \times N_R \times N_M \propto R^6$. However, since we apply FFT and the cross-correlation theorem to calculate the correlation between circles of the patterns rotated relative to each other, the factor $N_\varphi^2$ is replaced by $N_\varphi \times \log N_\varphi$. Therefore, the calculation time of our algorithm scales with $R^5 \times \log R$. This scaling behaviour was numerically tested and confirmed on several molecules in the 10-100 kDa range, and is favourable to the $R^6$-

$R^8$ scaling of other orientation algorithms (Fung et al. 2009, Loh and Elser, 2009, Moths and Ourmazd, 2011).

**Results and discussion**

We have tested the algorithm on simulated diffraction patterns of several molecules of various sizes. First a large number of diffraction patterns of the same molecule in random orientations were calculated. Then the patterns were oriented according to the described algorithm. The success of the orientation was analyzed based on the orientation information saved for each pattern in the preparatory phase. Finally, the structure was solved from the resulting intensity distribution.

*Measurement simulation*

A large ($N_M$) number of synthetic diffraction patterns of the molecule with random orientations were calculated. First, $N_M$ random 3D orientation matrices with uniform distribution on the whole SO(3) space of rotations were generated (Morawiec, 2004) and applied to the standard setting of the molecule. Atomic coordinates were taken from the Protein Data Bank and atomic scattering factors were calculated from the Cromer–Mann coefficients (Cromer and Mann, 1968). The simulated diffraction patterns were calculated as an intensity distribution on a polar grid. Poisson noise was calculated from the average counts/pulse in each pixel of the pattern.

*Orientation of the diffraction patterns*

Here we show how the algorithm works on the example of a large (109 kDa) protein molecule, the periplasmic nitrate reductase (Coelho et al. 2011) (*NapAB*), selected from the Protein Data Bank (PDB ID: 3ML1). First we generated 100,000 diffraction patterns of the molecule in different orientations. These orientations were saved for an analysis of the results only. The important parameters of the simulated patterns are listed in Table 1. We found that the very low intensity part of the patterns at high scattering angle was not useful for the determination of orientations. Therefore, data was truncated at $\vartheta'_{max} = 24°$ scattering angle, but the high resolution part of the pattern was kept for the purpose of structural reconstruction. We removed the average radial dependence from the patterns, and slightly smoothed them.

Starting from a random intensity distribution, convergence was reached after 9-15 iterations, depending on the random numbers. A single iteration takes 25 minutes on a workstation equipped with graphics processors. (All calculations were carried out on a workstation with two Intel Xeon X5680 processors and four Nvidia GTX 480 graphics cards.) We note here that in all cases when convergence was reached, less than 20 iterations was necessary, independent of the parameters used (size of the molecule, number, size and resolution of the patterns, noise level, etc.). For our example of the *NapAB* protein molecule, we show on Fig. 2 how the distribution of the correlation maxima changes as the iteration converges.

Since in this simulated case we know the true orientations of the diffraction patterns, we can calculate the accuracy of the orientation after convergence. The misorientation angle $\delta$ between orientations $\mathbf{R}_1$ and $\mathbf{R}_2$ is defined as the angle of rotation moving one orientation into the other: $\delta = \arccos((\text{Tr}(\mathbf{R}_1^T \mathbf{R}_2)-1)/2)$. This is a natural metric in the SO(3) rotational group (Morawiec, 2004). When convergence was reached, the average angular error is 1.1°, and all misorientations are smaller than 2.8°. The distribution of the misorientation angles is shown as empty bars on Fig. 3. After refining the orientation angles, the average angular error is reduced to 0.3° (full bars on Fig. 3).

*Structure reconstruction*

The real test of the method is whether one could reconstruct the electron distribution of the molecule from the oriented diffraction patterns. For this, we first constructed a 3D intensity distribution from the original noisy patterns with the refined orientations. Then we solved the phase problem using Fienup's hybrid input-output algorithm (Fienup, 1982) with parameter $\beta = 0.8$ to reconstruct the electron density in real space. A fixed spherical support with a diameter slightly larger than the molecule was used. For *NapAB*, the spacing of the density grid was 0.95 Å and the radius of the support was 46 Å. The resulting electron density is shown on Fig. 4 together with a ball-and-stick model of the original molecule. Since the electron density is reconstructed in a random orientation, the ball-and-stick model of the molecule was rotated and shifted with the help of the *Chimera* program (Pettersen et al. 2004). The agreement is almost perfect; there is no atom outside the electron density surface (set at 10% of the maximum density).

**Conclusions**

Here we demonstrated on the example of a large protein molecule, how a consistent 3D data set can be assembled from diffraction patterns of random orientations and the structure can be solved to atomic resolution. Our method is computationally much more efficient than other existing algorithms (Fung et al., 2009, Loh and Elser, 2009, Shneerson et al., 2008, Bortel and Tegze, 2011) even for relatively small *R* values and this advantage increases when the molecule is larger. This makes feasible to solve the structure of much larger objects with high resolution than it was possible before, which is crucial for the success of single molecule imaging experiments. In forthcoming publications we will show, that the method works also for symmetric objects, can find the symmetry elements without any prior knowledge, and can also select the diffraction patterns of one molecule out of a mixture.

**Acknowledgements** We are grateful to G. Oszlányi and G. Faigel for discussions and Z. Jurek and Á. Cserkaszky for technical help with the computer system and GPU programming. This work was supported by the Hungarian OTKA grants K81348 and K67980, and by the Bolyai János Scholarship of the Hungarian Academy of Sciences to GB.

**Figures**

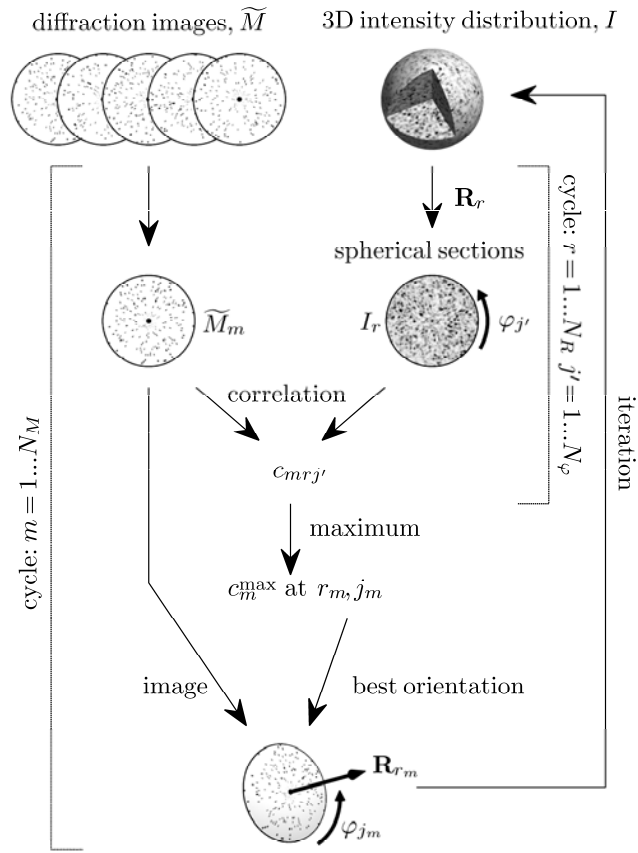

Figure 1. Orientation algorithm. The iteration begins with a random 3D distribution and a new intensity distribution is constructed in each step from all diffraction patterns rotated to their best-fitting orientations.

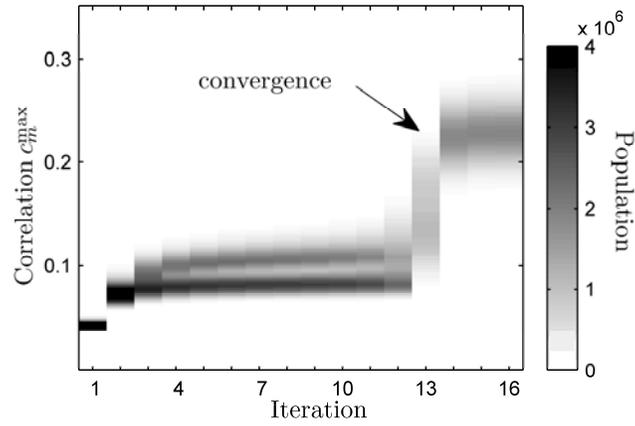

Figure 2. Evolution of the distribution of correlation maxima for the *NapAB* protein molecule. The sudden increase in the correlation indicates that the true orientations are found.

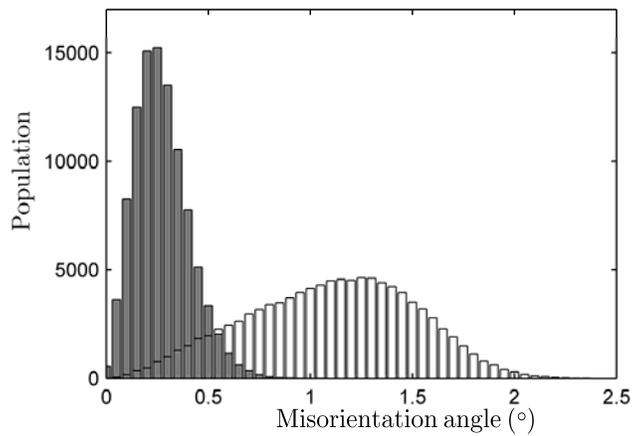

Figure 3. Distribution of the misorientation angle before (empty bars) and after (full bars) refinement of the orientation angles.

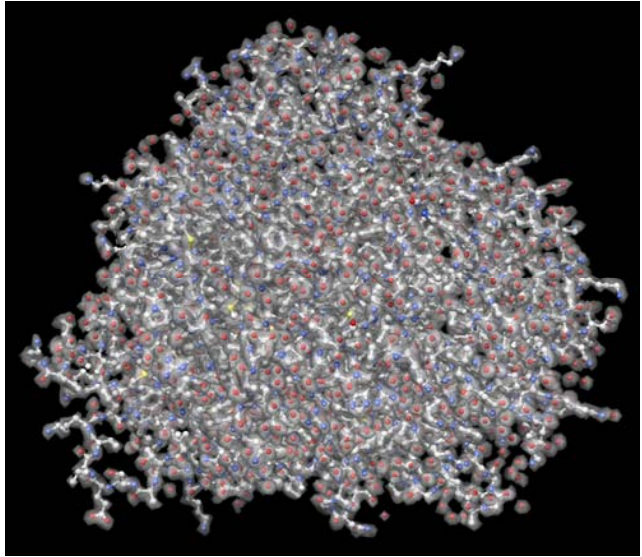

Figure 4. Electron density of the *NapAB* protein molecule calculated from 100,000 simulated noisy diffraction patterns of the randomly oriented molecule overlaid on its ball-and-stick model. The model is shifted and rotated to the best fitting position.

**Tables**

Table 1. Parameters of simulated diffraction patterns and grids

| | |
|---|---|
| Diameter of the molecule, $D$ | 91 Å |
| Wavelength of the radiation, $\lambda$ | 1 Å |
| Pulse fluence | $4 \cdot 10^{12}$ photons/(0.1μm×0.1μm) |
| Crystallographic resolution, $d$ | 1.9 Å |
| Minimum scattering angle, $\vartheta_{min}$ | 0.375° |
| Maximum scattering angle (used in reconstruction), $\vartheta_{max}$ | 30° |
| Maximum scattering angle (used in orientation), $\vartheta'_{max}$ | 24° |
| Pattern pixel size (polar × azimuthal), $\Delta\vartheta \times \Delta\varphi$ | 0.375°× 0.75° |
| Average total photon counts in pattern | 3300 |
| Average photon counts in outer pixels | 0.0131 |
| Average photon counts in outer Shannon–Nyquist pixels (Huldt et al., 2003) of size $(\lambda/2D)^2$ | 0.0092 |
| Number of patterns, $N_M$ | 100,000 |
| Number ($N_R$) and spacing of grid points in the $(\Theta,\Phi)$ orientation subspace | 5292 ~3° |
| Number of grid points (voxels) in the reciprocal space (used in orientation), $N_I$ | $129^3$ = 2,146,689 |